\documentclass[prd,twocolumn,superscriptaddress,nofootinbib,showpacs]{revtex4}

\usepackage{graphicx,amsmath,amsfonts,amssymb,revsymb,dcolumn,epsfig,bm}

\def\be{\begin{equation}}
\def\ee{\end{equation}}
\def\bea{\begin{eqnarray}}
\def\eea{\end{eqnarray}}

\begin{document}
\title{What can the detection of a single pair of circles-in-the-sky
tell us about the geometry and topology of the Universe?}

\author{B. Mota}
\affiliation{Universidade Federal do Rio de Janeiro,
NACO - CCS - Av. Brigadeiro Trompowski s/n°\\
21941-590 Rio de Janeiro -- RJ, Brazil}

\author{M.J. Rebou\c{c}as}
\affiliation{Centro Brasileiro de Pesquisas F\'{\i}sicas,
Rua Dr.\ Xavier Sigaud 150 \\
22290-180 Rio de Janeiro -- RJ, Brazil}

\author{R. Tavakol}
\affiliation{Astronomy Unit, School of Mathematical Sciences,
Queen Mary University of London\\ Mile End Road, London, E1 4NS, UK}

\date{\today}

\begin{abstract}

In a Universe with a detectable nontrivial spatial topology the
last scattering surface contains pairs of matching circles with
the same distribution of temperature fluctuations --- the so-called
circles-in-the-sky. Searches undertaken for nearly antipodal pairs of
such circles in cosmic microwave background maps have so far been unsuccessful.
Previously we had shown that  the negative outcome of such searches, if confirmed,
should in principle be sufficient to exclude a detectable non-trivial
spatial topology for most observers in very nearly flat
($0<\mid\Omega_{\text{tot}}-1\mid \lesssim10^{-5}$) (curved) universes.
More recently, however, we have shown that this picture is fundamentally changed
if the universe turns out to be {\it exactly} flat. In this case
there are many potential pairs of circles with large deviations from
antipodicity that have not yet been probed by existing searches.
Here we study under what conditions the detection of a single
pair of circles-in-the-sky can be used to uniquely specify the
topology and the geometry of the spatial section of the Universe.
We show that from the detection of a \emph{single} pair of matching
circles one can infer whether the spatial geometry is flat or not, and if
so we show how to determine the topology (apart from
one case) of the Universe  using this information.
An important additional outcome of our results is that
the dimensionality of the circles-in-the-sky parameter space that needs
to be spanned in searches for matching pairs of circles is reduced from
six to five degrees of freedom, with a significant reduction in the
necessary computational time.

\end{abstract}


\maketitle

\section{Introduction} \label{Sec0}
The determination of the shape or topology of the Universe constitutes one of
the fundamental open questions in cosmology%
\footnote{In line with the usage in the literature,
by topology of the Universe we mean the topology of its spatial sections.}
(see, e.g., the reviews~\cite{CosmTopReviews}).
An important observable signature of a universe with a detectable nontrivial
spatial topology is the presence in the cosmic microwave background (CMB) sky
of pairs of matching circles --- the so-called 
circles-in-the-sky --- with identical  distributions (up to a phase) of
temperature fluctuations%
~\cite{CSS1998}.
These pairs of circles can in turn be specified in terms of their associated
parameters, as for example their radii, relative phases, and deviations from
antipodicity.%
\footnote{This deviation refers to pair of circles whose centers are antipodal points
on the CMB sphere, which are known as back-to-back or antipodal circles-in-the-sky.}
The topology of a compact $3-$manifold can be uniquely specified
in terms of its  holonomy group. Furthermore, each distinct pair of circles
can be identified with a distinct element of this group.

Given the recent accumulation of high resolution
observations that indicates that the Universe is flat or nearly flat, a great
deal of effort has gone into studying the observable signatures
of cosmic topology in flat and nearly flat universes
(see, for example, Refs.~\cite{Some_Topol_Refs}).
An important outcome of our previous works has been that for detectable
pairs of circles in a very nearly flat compact universe, the deviations from
antipodicity will be small for most observers~\cite{Mota-etal-04,Mota-etal-08}.
This result, together with the  negative result of recent searches
for antipodal and nearly antipodal circles~\cite{Cornish-etal03,Key-et-al07},
had been taken to be sufficient to exclude a detectable nontrivial topology
for most observers~\cite{Mota-etal-08}.
More recently, we considered the case of an {\it exactly} flat universe.
By making a detailed study of all the compact orientable flat
manifolds~\cite{Mota-etal-10}, we found that the deviation from antipodicity in
some compact orientable flat universes can be larger than $10^\circ$, i.e. outside
the parameter ranges covered in the searches of Refs.~\cite{Cornish-etal03,Key-et-al07}.
This result has the important consequence that if the Universe is in fact flat
then the searches undertaken so far, which have confined themselves
to small deviations from antipodicity, would not be sufficient to rule out
the possibility of a nontrivial cosmic topology.%
\footnote{This important difference  comes ultimately from the very fact
that flat, spherical and hyperbolic classes of manifolds are topologically disjoint
in that their discrete topological invariants are different, and do not transform
smoothly between the classes. In other words, the nonexistence of a smooth limit
results comes from the fact that the classes of multiply-connected manifolds
associated with positive, negative and zero spatial curvatures are not only
topologically inequivalent, but also very dissimilar to one another.
In particular, while the non-flat manifolds are rigid (their compactification
lengths being topological invariants in the natural scale provided by
the curvature radius), the flat are not, and are therefore free to have
different sizes.}

This possibility  raises in turn the very interesting question  of the extent to
which the geometry and topology of the spatial sections of the Universe can be determined,
if a single pair of circles is detected in any future search of CMB data.
Here we make a systematic study of this question and show that, interestingly,
from the detection of a single pair of matching circles-in-the-sky
one can infer whether the spatial geometry is flat, and if
so what is the specific compact orientable flat topology
of the Universe (apart from one case).

The structure of the paper is as follows. In Sec.~\ref{SecII}
we derive the relations between the circles-in-the sky and
the holonomy parameters.
In Sec.~\ref{SecIII} we discuss the extent to which the geometry and
topology of a universe can be specified, given the detection of a single pair
of circles and its  related parameters.
In Sec.~\ref{SecIV} we present our final remarks and conclusions.

\section{Relating the circles-in-the-sky and holonomy parameters} \label{SecII}

Let us begin by recalling that homogeneous and isotropic spatial sections $M$ of
flat universes are often assumed to be the simply connected euclidian
$3$-manifold $\mathbb{R}^{3}$. However, they can also be multiply
connected quotient $3$-manifolds (which we assume to be compact and orientable)
of the form  $M=\mathbb{R}^{3}/\Gamma$, where $\mathbb{R}^{3}$ is the covering space,
and $\Gamma$ is a discrete and fixed point-free group of
isometries of $\mathbb{R}^{3}$ called the holonomy group~\cite{Thurston}.
Throughout this paper a generic element of the group $\Gamma$ is called holonomy
and denoted by $\gamma$.

In a Universe with a detectable non-trivial topology~\cite{TopDetec}
each pair of circles is identified with one element
of the corresponding holonomy group.
An important feature of compact flat manifolds is that any holonomy
$\gamma$ of an orientable Euclidean $3-$space can always
be expressed as a so-called screw motion (in the covering space),
consisting of a combination of a rotation (twist)
$R(\alpha,\mathbf{\widehat{u}})$ by an angle $\alpha$ around an axis%
\footnote{The choice of axes to describe a screw motion is not unique,
but one can always find, and we shall henceforth assume, a rotation axis
parallel to the direction of translation.}
of rotation $\mathbf{\widehat{u}}$, followed by a translation along 
a vector $\mathbf{L} = L \mathbf{\widehat{v}}$, say.
The action of $\gamma$ on any point $\mathbf{p}$ in the covering manifold is then
given by $\mathbf{p} \rightarrow R_{}^{}\,\mathbf{p} + \mathbf{L}$.
When there is no rotational part in the screw motion, i.e., when
$\alpha=0$, the holonomy reduces to a pure translation, and its action
is exactly the same at every point in space. In this case, the distance
between $\mathbf{p}$ and its image by the holonomy $\gamma$,
$\,\ell_{\gamma} = \,\mid\gamma\,\mathbf{p}-\mathbf{p}\mid\, = L$, is the
same everywhere.
On the other hand, for a general screw motion with $\alpha \neq 0$, $\ell_{\gamma}$
depends on the location of $\mathbf{p}$, and in particular on the distance $r$
between $\mathbf{p}$ and the axis of rotation.
Compact  flat manifolds are not rigid,
in the sense that topologically equivalent flat quotient manifolds,
defined by a given holonomy group $\Gamma$, can have different sizes,
and therefore their  compactification lengths are not fixed.
However, since the holonomy group must be a discrete and freely-acting
group of isometries of the covering space, the twist parameter $\alpha$
can only assume one of the following discrete
values~\cite{Adams-Shapiro01,Cipra02,Riazuelo-etal04}:
\be
\label{alphan}
\alpha = \frac{2\,\pi}{n}\,, \quad \textrm{with} \quad n=1,2,3,4,6\,.
\ee
With combinations of these holonomies one can construct the $6$ possible classes
of flat compact orientable manifolds. In all but one case ($E_6$), one can obtain
the full set of holonomies from combinations of two translational ($\alpha=0^\circ$)
and one screw-motion. Thus, one has the following five manifolds:
$E_1$ ($3$--torus, $\alpha=0^\circ$),
$E_2$ (half turn space, $n=2$, $\alpha=180^\circ$),
$E_3$ (one quarter turn space, $n=4$, $\alpha=90^\circ$),
$E_4$ (one third turn space, $n=3$, $\alpha=120^\circ$) and
$E_5$ (one sixth turn space, $n=6$, $\alpha=60^\circ $).
The sixth manifold, the Hantzsche-Wendt space, has as generators of the holonomy
group three screw motions, all with $n=2$ and $\alpha=180^\circ$.

\begin{figure}[tb!]
\begin{center}
\includegraphics[width=8.7cm,height=8.7cm,angle=0]{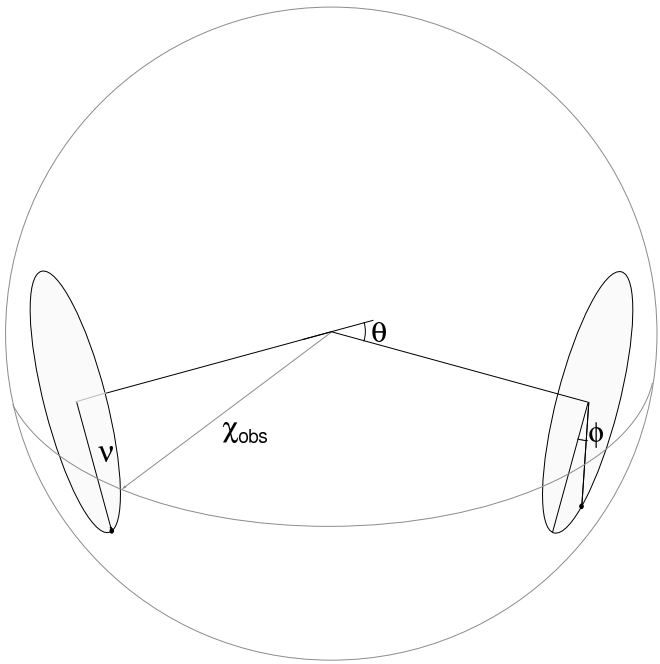}
\caption{\label{Fig1} This figure depicts the circles at the intersection
of the last scattering surface with its images.
The parameter $\theta$ is a measure of the
deviation of the circles from being antipodal, and $\phi$ measures the
relative phase between the circles of radius $\nu$.}
\end{center}
\end{figure}

The matching circles associated with the holonomy
$\gamma = \left( R(\alpha,\mathbf{\widehat{u}}), \mathbf{L} \right)$
are situated along the intersections of the sphere of last scattering with
its images under the holonomies $\gamma$ and $\gamma^{-1}$.
Each such pair of circles can be characterized by six angles, three of which are
({\bf i}) the deviation from antipodicity, $\theta$ ($0 \leq \theta \leq \pi$),
i.e. the complement of the angle between the centers of the matched circles,
({\bf ii}) the angular radius of the circles, $\nu$ ($0 \leq \nu \leq \pi/2$),
and ({\bf iii}) the phase shift $\phi$ ($0 \leq \phi \leq \pi$), which is
the phase angle that measures the shift in the identical distribution of CMB
temperature fluctuations of the two circles (see Fig.~\ref{Fig1} for details).
The remaining three angles give the position of the  center of the first circle
and the relative orientation of the second circle.

Now, from  Fig.~\ref{Fig1} one can show that for a pair of circles
produced by the holonomy $\gamma$, the angular radius $\nu$ is related
to the radius of the last scattering surface $\chi_{obs}$
and the  distance $\ell_\gamma$ between the observer's position and
its nearest image (the length of the shortest closed geodesic) through
\be
\label{sinalpha}
\sin \nu = \sqrt{1- \left (\frac{\ell_\gamma}{2\chi_{obs}} \right )^2} \,\,.
\ee

It follows directly from this equation that the only holonomies which generate
detectable pairs of circles are those for which $\ell_\gamma$ is short enough
so that $ \ell_\gamma< 2 \chi_{obs}$.
Furthermore, for these holonomies the shorter the distance $\ell_\gamma$ the
larger will be the circle radius $\nu$ and thus the holonomy with the shortest
distance between the observer and its image, $\ell_\gamma$, will produce the
matching circles with the largest radii.

The detection of a pair of matching circles would of course imply that the Universe
has multiply-connected spatial sections; but beyond that, we want to know how
much more information can be gleaned from such a detection.
We shall make explicit how and to what extent  the detection of a circle-in-the sky 
could be used to constrain the geometry and topology of the spatial section
of the Universe, as well as our position in it. To do so, we must first relate
the parameters specifying the circles to the elements of the associated holonomy.

To this end, let us assume that the spatial sections of the Universe
have a flat topology so that all holonomies take
the form of screw motions.
In addition, suppose that we have detected a pair of circles of radius $\nu$ with
parameters $\theta$, $\phi$, which is assumed to be the
most readily detectable pair of circles, i.e. the pair associated
with the shortest closed geodesic which contains the observer's position
or equivalently pairs with the circles with largest radius.%
\footnote{Clearly one might also detect other pairs of circles which are not associated
with the first neighboring copies of the CMB sphere. But in such a case these pairs
of circles corresponding to neighboring CMB sphere copies, i.e. associated to the shortest
closed geodesic, will also be detected, and the pairs could be distinguished by their
radii.}

The important question then is, given the parameters $\theta$, $\phi$ and $\nu$
of a single pair of circles, how to {\it uniquely} determine the parameters of
the corresponding screw motion, namely the angle $\alpha$, the compactification
length $L$, and the distance $r$ of observer to the axis of rotation?

\begin{figure}[tb!]
\begin{center}
\includegraphics[width=9cm,height=9cm,angle=0]{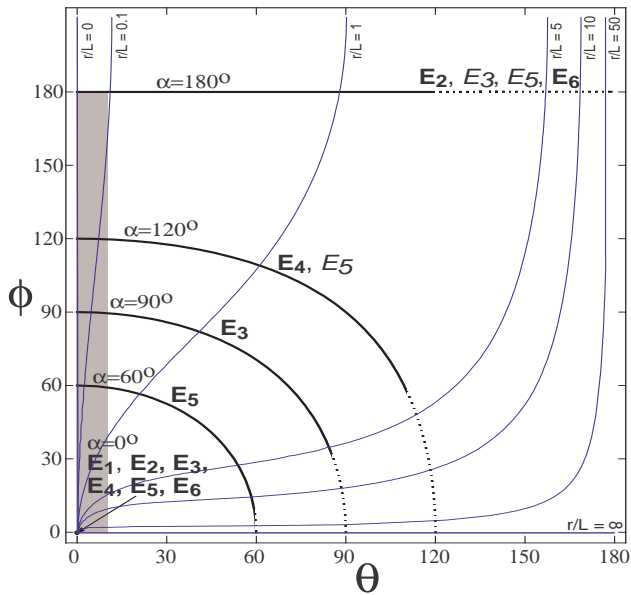}
\caption{\label{Fig2} The allowed combinations of the circles-in-the-sky parameters
$\theta$ (the deviation from antipodicity) and $\phi$ (the relative phase between circles)
for compact orientable flat manifolds. Each thick contour line corresponds to one allowed
class of holonomies, which take the form of a screw motion with twist parameter $\alpha$
given in Eq.~(\ref{alphan}). The different orientable compact flat manifolds $E_i\,$ ($i=1 \cdots 6$)
that include each holonomy class are indicated; those for which this holonomy may generate
the most readily detectable circle pair are in bold. The dotted part in each countour line
corresponds to configurations where the circle pairs with the largest radii are always
translational, and thus antipodal. In all cases, the compactification length $L$ is
not fixed, but the precise combination of $\theta$  and $\phi$ will depend on the
position of the observer, as depicted in the light contours corresponding to different
distances $r$ from the axis of the screw motion (in units of $L$).
The shaded area corresponds to the parameter values that have been probed in the recent
searches for antipodal and nearly antipodal circles in CMB
maps~\cite{Cornish-etal03,Key-et-al07}. } 
\end{center}
\end{figure}

To answer this question, we need to derive the relation between the
parameters specifying the holonomy and those corresponding to the circles.
To begin with the screw motion twist angle,
$\alpha$ can be uniquely determined by the phase shift $\phi$ and the
deviation from antipodicity $\theta$ through the relation
\be
\label{alpha}
\cos \alpha =  \frac{(\cos \phi+1)(\cos \theta+1)}{2}-1\,,
\ee
obtained by inverting the expression Eq.~(12) of Ref.~\cite{German03}.
Clearly, for a given value of $\alpha$ and $\theta$, there is one and only one
possible value of $\phi$. Conversely, the determination of both $\theta$
and $\phi$ specifies $\alpha$.
In this way, from Eq.~(\ref{alphan}) one has that the compact orientable flat manifolds
$E_i$ ($i = 1 \cdots 6$) define contour curves in the($\theta$--$\phi$) plane,
which are the loci of values of the parameters ($\theta,\phi$) allowed for the
circles-in-the-sky of flat universes whose spatial section is one of the associated flat
$3-$manifolds. The thick lines in Fig.~\ref{Fig2} indicate
these contour curves (see below for more details about this figure).

One can also show, after some algebra, that the compactification length $L$ and the
distance  of the observer $r$ to the axis of the screw motion can be written as
\be
\label{L}
L = 2\,\chi^{}_{obs}\,\cos \nu\,\,\sqrt{\frac{\cos\theta -\cos\alpha}{1 - \cos \alpha}} \,,
\ee
and
\be
\label{r}
r = \sqrt{2}\,\chi^{}_{obs}\,\cos\nu\,\, \frac{\sqrt{1-\cos \theta}}{1 - \cos \alpha} \,.
\ee
respectively. Thus, given the parameters of the circles, ($\theta, \phi, \nu$),
and the radius of the last scattering surface, $\chi_{obs}$,
one can obtain the parameters for the corresponding holonomy ($\alpha, L$)
and the distance $r$ of the observer  to the rotation axis (henceforth, the observer's position).
Clearly, from Eqs.~(\ref{r}), (\ref{L}) and~(\ref{alpha}) one obtains the locus of values
of the parameters ($\theta,\phi$) such that the ratio $r/L$ is constant.
This locus defines another family of contour curves in the ($\theta$--$\phi$) plane,
depicted in Fig.~\ref{Fig2} as thin traversing curves for different values of ratio $r/L$.
Different combinations of $\theta$ and $\phi$ along each contour correspond to
different values of $r$ (in units of the compactification length $L$).
Figure~\ref{Fig2} shows that for observers situated along the screw motion
axis the resulting circles are antipodal ($\theta =0$), with a relative
phase $\phi$ given by the twist angle $\alpha$ [cf.\ Eq.~(\ref{alpha})].
For all observers  the deviation from antipodicity
$\theta$ becomes larger as the phase $\phi$ decreases.
In the limit where the observer is infinitely distant from the axis,
$\phi$ becomes zero and $\theta$ becomes equal to $\alpha$ [see Eq.~(\ref{alpha})].

\section{Circles-in-the-sky in compact orientable flat universes} \label{SecIII}

Now let us assume observations have detected a single pair of circles of radius
$\nu$ with the corresponding parameters  ($\theta, \phi$).
As we mentioned above, in compact orientable flat manifolds the screw motion angle
$\alpha$ is restricted to the values  given by Eq. (\ref{alphan}).
Also, for each of the allowed values, Eq.~(\ref{alpha}) becomes a finite set of one-to-one
relations between $\theta$ and $\phi$,  as is shown in  Fig.~\ref{Fig2}.
Thus, the detection of a pair of circles will allow the knowledge of the parameters
$\theta$ and $\phi$ to readily determine whether  or not the geometry of the spatial section of
the Universe is Euclidean (flat), by checking whether the observed values
of $\theta$ and $\phi$ lie on one of the contours (thick curves) shown in Fig.~\ref{Fig2}.
If they do not (after taking into account observational uncertainties)
then this is a clear indication that the geometry is non-Euclidean.
Conversely, if the parameters $\theta$ and $\phi$ of the detected pair lie, within
observational uncertainty limits,  on one of these thick contours curves, we would
conclude that the underlying spatial geometry is most likely Euclidean (flat).%
\footnote{The values along the contours are in fact compatible with some positions
of the observer for certain curved manifolds, but the full set of possible
combination of the latter densely span the $\theta$ -- $\phi$ plane, and thus
the set of values corresponding exactly to the contour curves correspond to
a zero-measure set of observers for each potential non-Euclidean  manifold.}

Now if it is found that the geometry is indeed Euclidean, we then wish to establish 
to what extent the topology of the spatial section of the Universe can be determined,
given such a detected  pair of circles.
The list of possibilities of compact orientable flat manifolds is summarized in
Table~\ref{table:theta-max}. We indicate in each case the
maximum deviations from antipodicity of the circles-in-the-sky
for which a non-translational holonomy may generate the most readily
detectable pair of circles (see Ref.~\cite{Mota-etal-10} for details).

\begin{table}[ht!]
\begin{tabular}{*4{c}}  
 \hline\hline
Symbol   & Manifold                     &    $n$   & \hspace{2mm} $\theta_{\text{max}}$  \\
\hline
$E_1$    &   three-torus                & 1,1,1  & \hspace{2mm} $0^{\circ}$    \\
$E_2$    & half turn space              & 1,1,2  & \hspace{2mm} $120^{\circ}$  \\
$E_3$    & quarter turn space           & 1,1,4  & \hspace{2mm} $86^{\circ}$   \\
$E_4$    & third turn space             & 1,1,3  & \hspace{2mm} $109^{\circ}$  \\
$E_5$    & sixth turn space             & 1,1,6  & \hspace{2mm} $59^{\circ}$   \\
$E_6$    & Hantzsche-Wendt space        & 2,2,2   & \hspace{2mm} $120^{\circ}$  \\
\hline\hline
\end{tabular}
\caption{Multiply-connected flat orientable manifolds and the maximum deviation from
antipodicity of the circles-in-the-sky for each manifold for which a non-translational
holonomy may generate the most readily detectable pair of circles(i.e., with the largest
radii). In all cases the screw motion twist parameters can only take certain values
of the form $\alpha=2\pi /n $; the values of $n$ for the holonomy group generators
are also indicated. Note that $n=1$ corresponds to a translation.} \label{table:theta-max}
\end{table}

Table~\ref{table:theta-max} and Fig.~\ref{Fig2} show that
each of the $6$ possible classes of holonomies corresponds to
the different values of $\alpha$ [cf.\ Eq.~(\ref{alphan})], which in turn
belong to the holonomy group of one or more of the compact
orientable flat manifolds.

At first sight, this seems to indicate one-to-many correspondence between
the values of the twist angle  $\alpha$ [obtained  from the
detected angular parameters of circles ($\theta,\phi$)] and the list of
flat orientable compact manifolds $E_i$ ($i=1...6$) with their associated
holonomies.

However, as our previous work~\cite{Mota-etal-10} details, some elements
of a holonomy group can never produce the pair of circles with the shortest
distance between the observer and its image, no matter where in the
manifold the observer happens to be.
For instance, although a screw motion with a twist of $120^\circ$ exists in
the holonomy group of $E_5$, being the square of its non-translational
generator with a twist of $60^\circ$, it can be shown that the distance
between any point and its image by the former holonomy is always larger
than the corresponding distance by the latter.
Thus, if a pair of circles associated with a twist of $120^\circ$ is detected,
it means that either the spatial section of the Universe is $E_4$, or it is $E_5$ but
there is another pair of matching circles, of larger radii (and thus more
readily detectable), the position of which can be worked out from the position
of the centers of the circles and from Eqs.~(\ref{alpha}), (\ref{L}) and (\ref{r}).%
\footnote{Clearly, we are ignoring the possibility of having such circles hidden
by galactic contamination.}
If, on the other hand, no other pair is detected in a full-sky search,
then merely the presence of this pair of circles is sufficient to ensure that the
spatial section has an $E_4$ topology.
In this way, a non-antipodal circle pair detection with parameters on any
contour line in Fig.~\ref{Fig2},
is compatible with only one flat compact orientable manifold, apart from one case:
a circle pair with phase shift of $\phi = 180^\circ$ for any value of $\theta$
(implying a screw motion with a twist of $\alpha = 180^\circ$) may be the most
readily detectable circle pair in either $E_2$ and $E_6$
(and can also be present in $E_3$ and $E_5$, albeit necessarily
alongside some other circle pair with larger radii).
Thus, the observation of a single pair of non-antipodal circles removes the
ambiguities in determining the spatial topology and, except for the degeneracy
between $E_2$ and $E_6$, it allows one to fully reconstruct the entire holonomy
group (apart from at most two compactification lengths associated with the group
generators not directly observed), as well as to determine observer's position
within the compact manifold.

This leaves out the antipodal case given by $\theta= 0 = \phi$, where
the circles are back-to-back with no phase difference. As can be
seen from Fig.~\ref{Fig2}, this is a totally degenerate case,
since for any flat $3-$manifold a suitable choice of position
and compactification lengths guarantee that a translation
generates the only detectable circle pair. Note, however, that
this case has been in principle ruled out by the searches that
have already taken place.

In closing, we note that our results can also be useful to devise
search strategies for circles-in-the-sky in CMB maps that reduce
the size of the parameter space that needs to be numerically searched.
This is potentially significant because a pair of matching circles
in the celestial sphere is defined by six angular parameters.
A six-parameter search in high-resolution maps to be produced from the ongoing
Planck mission~\cite{Planck-Collab} would be prohibitively time-consuming.
By combining our previous result according to which the deviation from antipodicity
is very likely to be small for circle pairs in very nearly flat universes,
with the connection shown here to exist between two of the circle pair
parameters in flat universes, our results suggest that future searches
should be confined to combinations of $\theta$ and $\phi$ given by Eq.~(\ref{alpha})
restricted to Eq.~(\ref{alphan}). Clearly, in order to carry out a new search
for nearly back-to-back pairs of circles, an additional full
range of $\phi$ values for small values of $\theta$ should
be undertaken.
In this way, one would effectively reduce the dimension of the
survey parameter space by one, ensuring a significant reduction
in computational time.

\section{Concluding Remarks} \label{SecIV}

The existence of pairs of correlated circles in CMB maps with the same
distribution of temperature fluctuations --- the so-called circles-in-the-sky ---
is a generic prediction in a Universe with a detectable nontrivial cosmic
topology, regardless of the background geometry.
Detecting such  circle pairs would  provide a measure of the value of their
corresponding parameters ($\theta, \phi, \nu)$).

We have made a detailed study of the extent to which the detection of a single
pair of circles in CMB maps can be used to determine the geometry and topology of the
spatial sections of the Universe.

We have shown that as long as we detect the pair of circles which is the most readily
detectable one, i.e. associated with the nearest (topological) copies of the
last scattering surface, then
the analysis of the corresponding circle parameters given in this paper
is sufficient to determine whether the geometry is Euclidean (flat) or curved,
within the observational uncertainties.
In the former case, we have also shown that we can fully determine the topology
of the spatial sections of the Universe, apart from one case.

Given the upcoming high resolution data from Planck~\cite{Planck-Collab},
it is conceivable that pairs of circles-in-the-sky
can be detected by more comprehensive searches than those so far
undertaken~\cite{Cornish-etal03,Key-et-al07}.
However, any detection is likely to be partial, in the sense that
it is very  unlikely that one would have access to a complete
set of generators of the holonomy group from the observable  circles-in-the-sky.
In spite of this, what we have shown in this paper is that even the most
partial detection, i.e. the observation of a single pair of circles, is in principle
sufficient to determine whether the spatial section of the Universe is  flat or curved.
Furthermore, if it is flat, we have shown how one can determine the topology of the
spatial section of the Universe from this minimal detection.

Finally our results have an important consequence for the future search for
circles-in-the-sky strategies. We had previously shown that searches so far
undertaken (if confirmed) would be sufficient to in principle exclude the
possibility of a very nearly flat universe with a non-trivial
topology~\cite{Mota-etal-04,Mota-etal-08}.
In the light of this result, our results in the present paper suggest
where in the parameter space the searches for circles-in-the-sky one
should concentrate on, thus significantly reducing their computational costs.

\begin{acknowledgments}
M.J. Rebou\c{c}as acknowledges the support of FAPERJ under a CNE E-26/101.556/2010 grant.
This work was also supported by Conselho Nacional de Desenvolvimento
Cient\'{\i}fico e Tecnol\'{o}gico (CNPq) - Brasil, under grant No. 475262/2010-7.
M.J. Rebou\c{c}as and R. Tavakol thank, respectively,  CNPq and
PCI-CBPF/MCT/CNPq for the grants under which this work was carried
out.
\end{acknowledgments}



\end{document}